# Континуальная модель одномерного биполярона Холстейна в ДНК

©2014 Каширина Н.И.[*,1], Лахно В.Д.[**,2]

[1]*Институт физики полупроводников им. В.Е Лашкарева, Национальная академия наук Украины, Киев, 03028, Украина*

[2]*Институт математических проблем биологии, Российская академия наук, Пущино, Московская область, 142290, Россия*

***Аннотация.*** В работе в континуальном приближении получен функционал 1D-биполярона Холстейна. Изучается влияние электронных корреляций, связанных с прямой зависимостью волновой функции системы от расстояния между электронами, на энергию связи биполярона. Для различных параметров рассматриваемой системы проведены расчёты энергии связи биполярона, получена фазовая диаграмма области устойчивости биполярона.

***Ключевые слова:*** *электрон-фононное взаимодействие, полярон Холстейна, биполярон.*

## ВВЕДЕНИЕ

Сверхпроводящее состояние в ДНК наблюдалось при низких температурах для хаотических нуклеотидных последовательностей ($\lambda$-ДНК) [1]. В связи с возможностью объяснения сверхпроводимости в ДНК на основании биполяронного механизма, значительный интерес представляет исследование различных моделей образования биполяронных состояний. Так как условие Бозе-конденсации биполяронного газа может быть затруднено неоднородностью нуклеотидного состава, в [2] предполагается, что низкая температура сверхпроводящего перехода $T \approx 1\,\text{K}$, наблюдавшаяся в [1], связана с хаотичным расположением нуклеотидных последовательностей. При переходе к однородным системам возможно наблюдение сверхпроводимости при более высоких температурах. Континуальная модель биполярона (БП) большого радиуса в применении к низкоразмерным системам, в частности к ДНК, интересна также в связи с тем, что в рамках данного рассмотрения эффективная масса БП, как правило, значительно ниже, чем значения, полученные для БП малого радиуса. Это может существенно повысить температуру перехода сверхпроводящее состояние. В связи с тем, что модель полярона Хаббарда–Холстейна допускает рассмотрение как полярона (П) малого радиуса, так и П большого радиуса, можно применить континуальный метод Холстейна, предложенный в [3], для получения функционала БП в континуальном приближении.

## ФУНКЦИОНАЛЫ ПОЛЯРОНА И БИПОЛЯРОНА ХОЛСТЕЙНА

В адиабатическом приближении волновые функции (ВФ) полярона $\Phi(x)$ [3] и биполярона $\Psi(x_1, x_2)$ удовлетворяют следующим уравнениям:

$$-J \frac{d^2}{dx^2} \Phi(x) - A \cdot u(x) \Phi(x) = W_P \Phi(x), \qquad (1)$$

$$-J \Delta_{12} \Psi_{12} - A(u_1 + u_2) \Psi_{12} + G \delta_{12} \Psi_{12} = W_B \Psi_{12}, \qquad (2)$$

где $\Psi_{12} = \Psi_{21} \equiv \Psi(x_1, x_2)$,

---

[*] kashirina@mail.ru
[**] lak@impb.psn.ru

$$\Delta_{12} = \frac{d^2}{dx_1^2} + \frac{d^2}{dx_2^2},$$

$u_1 \equiv u(x_1)$, $u_2 \equiv u(x_2)$ – операторы смещения в месте расположения соответственно 1-ой и 2-ой заряженной частицы, слагаемое $G\delta_{12}\Psi_{12}$ соответствует кулоновскому отталкиванию взаимодействующих частиц в 1D системе [4], $\delta_{12} = \delta(x_1 - x_2)$, $x_1, x_2$ – координаты 1 и 2 заряженной частицы. Величина $J$ согласно [3] может быть выражена через эффективную массу заряженной частицы $m^* = \hbar^2/2Ja^2$, где $a$ – расстояние между ближайшими соседями в линейной цепочке.

Функционалам П и БП соответствуют следующие выражения:

$$F_{P,B} = W_{P,B} + \frac{k}{2}\int u^2(x)dx, \tag{3}$$

где $k = M\omega_0^2$, $M$ – масса колеблющихся частиц в цепочке, $\omega_0$ – частота колебаний.

Приравняв нулю функциональную производную (3) по оператору смещения (адиабатическое приближение), выразим оператор смещения П и БП через электронную волновую функцию:

$$u_0(x) = \frac{A}{k}\Phi(x), \tag{4}$$

$$u_0(x) = \frac{2A}{k}\int |\Psi(x,x')|^2 dx'. \tag{5}$$

Подставляя (4) и (5) соответственно в (1) и (2), получим функционалы П и БП в следующем виде:

$$F_P = T_P + W_{P(\text{int})} + U_{P(ph)}, \tag{6}$$

$$F_B = T_B + W_{B(\text{int})} + U_{B(ph)} + V_C, \tag{7}$$

где

$$T_P = -J\int \Phi^*(r)\Delta\Phi(r)d\tau, \tag{8}$$

$$W_{P(\text{int})} = -\frac{A^2}{k}\int |\Phi(x)|^4 dx, \tag{9}$$

$$U_{P(ph)} = \frac{k}{2}\int u_0^2(r)d^3r = \frac{A^2}{2\cdot k}\int |\Phi(x)|^4 dx, \tag{10}$$

$$V_C = G\iint \Psi_{12}^*\delta_{12}\Psi_{12}dx_1 dx_2 = G\int |\Psi_{11}|^2 dx_1, \tag{11}$$

$$T_B = -J\int \Psi_{12}^*\Delta_{12}\Psi_{12}dx_1 dx_2, \tag{12}$$

$$W_{B(\text{int})} = -\frac{4A^2}{k}\int |\Psi_{12}|^2 |\Psi_{23}|^2 dx_1 dx_2 dx_3. \tag{13}$$

Подставляя выражения (8) – (13) в функционалы П (6) и БП (7) получим:

$$F_P = -J\int \Phi_1^*\Delta_1\Phi_1 dx_1 - \frac{A^2}{2k}\int |\Phi_1|^4 dx, \tag{14}$$

$$F_B = -J\iint \Psi_{12}^*\Delta_{12}\Psi_{12}dx_1 dx_2 - \frac{2A^2}{k}\iiint |\Psi_{12}^*|^2 |\Psi_{23}^*|^2 dx_1 dx_2 dx_3 + G\int |\Psi_{11}|^2 d^3x_1. \tag{15}$$

С учётом (4) уравнение для П Холстейна (1) в континуальном приближении имеет вид

$$-J\Delta\Phi(x) - \frac{A^2}{k}\Phi^3(x) = W_P\Phi(x). \tag{16}$$

Обозначив $E_{B(P)} = \min\{F_{B(P)}\}$, получим критерий устойчивости БП $2E_P - E_B = \Delta_B > 0$.

Как известно, уравнение (16) имеет точное аналитическое решение. Волновая функция П большого радиуса и соответствующая энергия основного самосогласованного состояния имеют вид

$$\Phi(x) = \alpha/\sqrt{2}\cosh(\alpha^2 x), \; E_P = -2J - \alpha^4 J/3, \tag{17}$$

где $\alpha^2 = \kappa/8J$, $\kappa = 2A^2/k$, $k = M\omega_0^2$, $M$ и $\omega_0$ – масса и частота колебаний сайта.

Для расчётов энергии связи БП могут использоваться различные наборы пробных ВФ, апробированных для расчётов энергии в 3D системах, например:

$$\Psi_{12} = N(1 + b(x_1 - x_2)^2)\exp(-ax_1^2 - ax_2^2), \tag{18a}$$

$$\Psi_{12} = N(x_1 - x_2)^2 \sum_i C_i \exp(-a_{i1}x_1^2 - 2a_{2i}x_1x_2 - a_{3i}x_2^2), \tag{18b}$$

$$\Psi_{12} = N\sum_i C_i \exp(-a_{1i}x_1^2 - 2a_{2i}x_1x_2 - a_{3i}x_2^2), \tag{18c}$$

где $N$ – нормировочный множитель, $a_1, a_2, a_3, a, b, C_i, a_{1i}, a_{2i}, a_{3i}$ – вариационные параметры.

## РАЗЛИЧНЫЕ ПРЕДСТАВЛЕНИЯ КУЛОНОВСКОГО ОТТАЛКИВАНИЯ В ОДНОМЕРНЫХ И КВАЗИОДНОМЕРНЫХ СИСТЕМАХ

Энергия связи континуального БП существенно зависит от учёта электронных корреляций. В одномерной системе выбор пробной электронной ВФ в виде $\Psi_{12} = f(x_1 - x_2)\Psi'_{12}$, где $f(x_1 - x_2)$ – функция координат электронов, выбранная таким образом, что для $x_1 = x_2$ $f(x_1 - x_2) = 0$, позволяет полностью исключить кулоновское отталкивание между заряженными частицами и, при удачном выборе пробной функции $\Psi'_{12}$, может значительно понизить энергию одноцентровой конфигурации 1D-полярона. Тем не менее, из физических соображений данное приближение кажется вполне оправданным для биполярона Холстейна малого радиуса, когда ВФ электронов практически полностью сконцентрирована на одном сайте. В том случае, когда речь идёт о кулоновском взаимодействии частиц большого радиуса, когда ВФ одномерной дырки (электрона) «размазана» по нескольким сайтам, представление кулоновского отталкивания дельта-образным потенциалом не кажется безупречной аппроксимацией. В том случае, когда данное взаимодействие не носит контактный характер (квазиодномерные системы), его фурье-компонента в $x$-направлении может быть представлена в виде [5]:

$$V_C(q_x) = \iint dq_z dq_y \frac{4\pi e^2}{q_x^2 + q_y^2 + q_z^2} \cdot \left[\int_0^\infty r dr |\chi(r)|^2 J_0(q_x r)\right]^2, \tag{19}$$

где $J_0(z)$ – функция Бесселя нулевого порядка, функция $\chi(r)$ зависит от выбранной аппроксимации кулоновского взаимодействия. Так, например, при выборе

$$\chi(r) = \left[\frac{2}{\pi}\right]^{1/2} \frac{1}{\langle r \rangle} \exp(-r/\langle r \rangle), \qquad (20)$$

величина $V_C(q_x)$ может быть вычислена аналитически, как функция параметра $\langle r \rangle$, представляющего собой величину, которая характеризует распределение электронной плотности в плоскости $\langle y-z \rangle$ вблизи направления оси $\langle x \rangle$. В отличие от двумерных систем, этот параметр не может полагаться равным нулю в связи с тем, что при $\langle r \rangle \to 0$ $V_C(q_x)$ стремится к бесконечности. Поэтому при такой аппроксимации кулоновского отталкивания параметр $\langle r \rangle$ играет роль дополнительного параметра теории, и вычисления энергии связи БП должны быть проведены для разных отличных от нуля величин параметра $\langle r \rangle$. Приближение, использовавшееся в [5] для 1D систем, близко к рассмотренному в [6] случаю БП в анизотропных кристаллах, когда переход к низкоразмерным 2D и 1D системам может быть получен численно изменением параметров анизотропии кристаллов. При этом кулоновское взаимодействие носит трехмерный характер.

## РЕЗУЛЬТАТЫ РАСЧЁТОВ

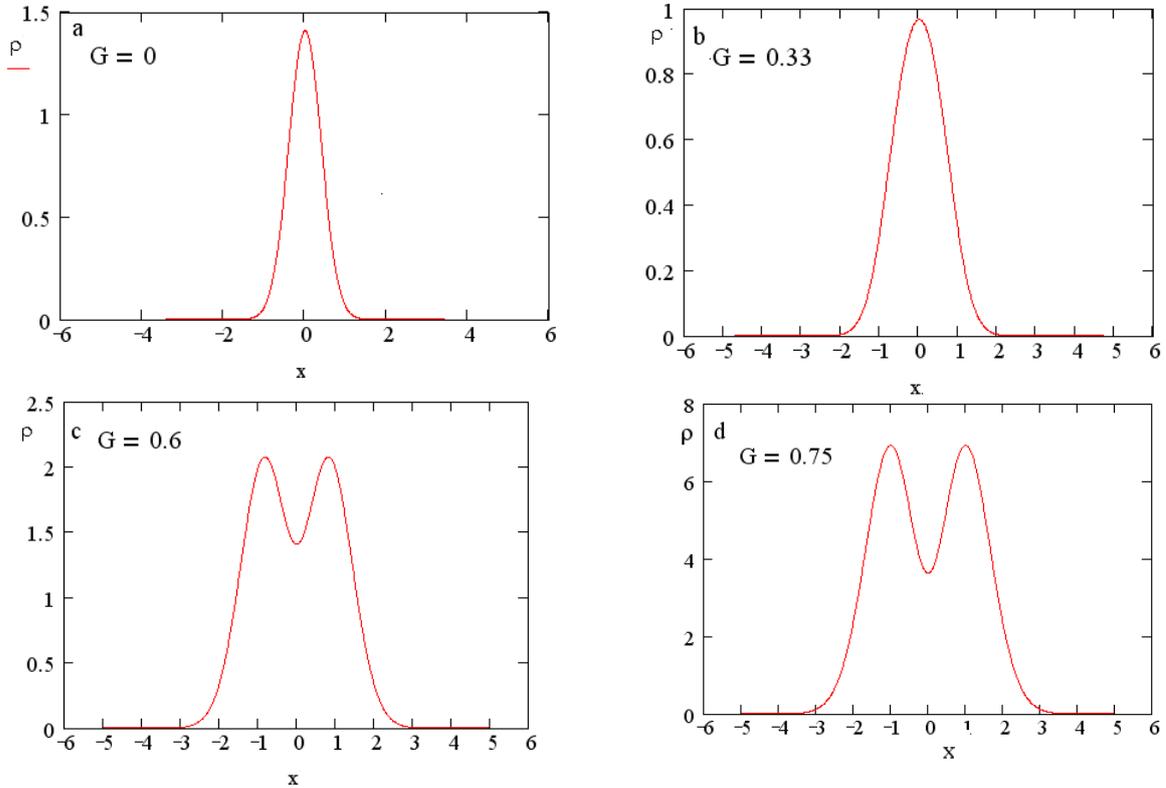

**Рис. 1**. Плотность заряда $\rho_1 \equiv \rho_1(x_1) = \int_{-\infty}^{\infty} |\Psi_{12}(x_1, x_2)|^2 dx_2$ для ВФ (18a) $\rho_i \equiv \rho_i(x_i)$.

На рис. 1 представлены графики плотности заряда как функции координаты электрона для ВФ (18a). По координате второго электрона проведено усреднение. Приведенные на рис. 1. графики хорошо согласуются с аналогичными зависимостями, приведенными в [2] для дискретной модели БП. Для удобства сравнения нами в качества примера построены зависимости $\rho(x)$ для тех же параметров, что и в работе [2], в которой исследовалась дискретная модель биполярона Холстейна–Хаббарда. Обратим внимание на то, что в связи с низкой размерностью исследуемой нами системы, наглядное отличие одноцентровой модели БП и двухцентровой модели, проявляющееся в трехмерном случае, теряется. Если в трехмерном случае одноцентровая модель БП – это сферически симметричное образование, в то время как двухцентровая модель соответствует аксиально-симметричному БП, то в одномерной системе учёт электронных корреляций (в нашем случае это прямая зависимость ВФ электронов от межэлектронного расстояния) также приводит к тому, что с ростом кулоновского отталкивания электронная плотность распределяется по двум областям, разделённым резким минимумом в начале координат. На рис. 1. величина $G$ выражена в эВ, $J = 0.084$ эВ, и $\kappa = 0.5267$ эВ, расстояние между сайтами принято за единицу.

**Таблица 1.** Удвоенная энергия основного самосогласованного состояния полярона $2E_p$, энергия биполярона $E_b$, энергия связи БП $\Delta$ в электрон-вольтах. Величины $\Delta$, $G$ и к выражены в электрон-вольтах. Пробная ВФ БП выбрана в виде (18a). Параметры $p$, $a$, $b$ – параметры минимизации П и БП функционала соответственно. ВФ полярона: $\Psi(x) = N\exp(-p^2 x^2)$, $N$ – нормировочный множитель.

| | $U$ | $-2E_p$ | $p$ | $-E_b$ | $a$ | $b$ | $\Delta = 2E_p - E_{Bp}$ |
|---|---|---|---|---|---|---|---|
| $\kappa = 0.1$ $J = 0.084$ | 0.1 | 0.001184 | 0.083956 | 0.001187 | 0.074043 | 0.026141 | 0.000003245 |
| | 0.075 | 0.001184 | 0.083956 | 0.001259 | 0.081677 | 0.019118 | 0.000075259 |
| | 0.05 | 0.001184 | 0.083956 | 0.001556 | 0.108316 | 9.4295E-3 | 0.000372757 |
| | 0.025 | 0.001184 | 0.083956 | 0.002734 | 0.136669 | 3.1097E-3 | 0.001550395 |
| | 0 | 0.001184 | 0.083956 | 0.004736 | 0.167913 | 0 | 0.003552566 |
| $\kappa = 0.2$ $J = 0.084$ | 0.2 | 0.004736 | 0.167913 | 0.004749 | 0.148087 | 0.104565 | 0.00001 |
| | 0.175 | 0.004736 | 0.167913 | 0.004861 | 0.15384 | 0.09098 | 0.000124 |
| | 0.15 | 0.004736 | 0.167913 | 0.005037 | 0.163354 | 0.076472 | 0.000301 |
| | 0.1 | 0.004736 | 0.1679135 | 0.006227 | 0.216633 | 0.037718 | 0.001491 |
| | 0 | 0.004736 | 0.1679135 | 0.018947 | 0.335827 | 0 | 0.014210 |
| $\kappa = 0.3$ $J = 0.084$ | 0.3 | 0.010657 | 0.251870 | 0.010686 | 0.222131 | 0.235272 | 0.000029 |
| | 0.275 | 0.010657 | 0.251870 | 0.010842 | 0.22746 | 0.215063 | 0.000185 |
| | 0.15 | 0.010657 | 0.251870 | 0.014012 | 0.324949 | 0.084865 | 0.003354 |
| | 0.1 | 0.010657 | 0.251870 | 0.020149 | 0.382178 | 0.040471 | 0.00949 |
| | 0 | 0.010657 | 0.251870 | 0.042630 | 0.503741 | 0 | 0.031973 |
| $\kappa = 0.4$ $J = 0.084$ | 0.4 | 0.018947 | 0.335827 | 0.018998 | 0.296175 | 0.418262 | 0.000029 |
| | 0.3 | 0.018947 | 0.335827 | 0.020151 | 0.326709 | 0.305890 | 0.000185 |
| | 0.2 | 0.018947 | 0.335827 | 0.024911 | 0.433266 | 0.150872 | 0.003354 |
| | 0.1 | 0.018947 | 0.335827 | 0.043753 | 0.546677 | 0.049756 | 0.00949 |
| | 0 | 0.018947 | 0.335827 | 0.075788 | 0.671654 | 0 | 0.031973 |
| $\kappa = 0.7$ $J = 0.084$ | 0.7 | 0.05802 | 0.587697 | 0.058184 | 0.518307 | 1.280917 | 0.000159003 |
| | 0.6 | 0.05802 | 0.587697 | 0.059796 | 0.542168 | 1.090127 | 0.001771705 |
| | 0.4 | 0.05802 | 0.587697 | 0.069058 | 0.685018 | 0.618775 | 0.011033118 |
| | 0.2 | 0.05802 | 0.587697 | 0.123031 | 0.928324 | 0.179210 | 0.065006196 |
| | 0 | 0.05802 | 0.587697 | 0.232100 | 1.175394 | 0 | 0.174075719 |
| $\kappa = 1$ $J = 0.084$ | 1 | 0.118418 | 0.839567 | 0.118743 | 0.740439 | 2.61413 | 0.000324495 |
| | 0.75 | 0.118418 | 0.839567 | 0.125944 | 0.816774 | 1.911829 | 0.007525872 |
| | 0.5 | 0.118418 | 0.839567 | 0.155694 | 1.083164 | 0.942954 | 0.037275732 |
| | 0.2 | 0.118418 | 0.839567 | 0.30750 | 1.425417 | 0.242007 | 0.189081373 |
| | 0 | 0.118418 | 0.839567 | 0.473675 | 1.679135 | 0 | 0.355256569 |

В таблице 1 приведены энергии основного состояния полярона и БП Холстейна, а также параметры минимизации для ВФ (18a).

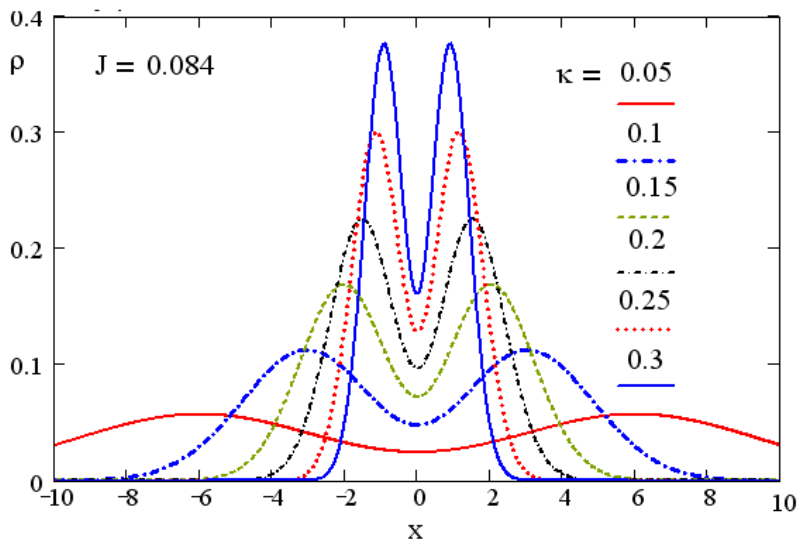

Рис. 2. Плотность заряда как функция координаты электрона для ВФ (18b) при различных параметрах электрон-фононного взаимодействия, определяемых параметром $\kappa = 2A^2/k$.

На рис. 2 приведены зависимости плотности электронных состояний ВФ биполярона, электронная часть которой выбрана в простейшем виде $\Psi_{12} = (x_1 - x_2)^2 \exp(-ax_1^2 - ax_2^2)$ (частный случай ВФ (18b)). При подобном выборе электронной ВФ, кулоновское взаимодействие, которое носит в рассматриваемой нами модели контактный характер, выпадает из рассмотрения. Тем не менее, ВФ, выбранная в таком виде, не приводит к связанному состоянию БП, несмотря на то, что её трехмерный аналог, рассматривавшийся, например, в работе, посвящённой расчётам энергии трехмерного одноцентрового БП [7], давал одно из наиболее низких связанных состояний континуального БП сильной связи. Более сложные многопараметрические функции (18b) требуют отдельного рассмотрения. Обратим внимание на то, что в общем случае более сложных многопараметрических ВФ (18b) и (18c) требуются дополнительные процедуры симметризации для синглетных состояний и антисимметризации ВФ для триплетных состояний двухэлектронных систем.

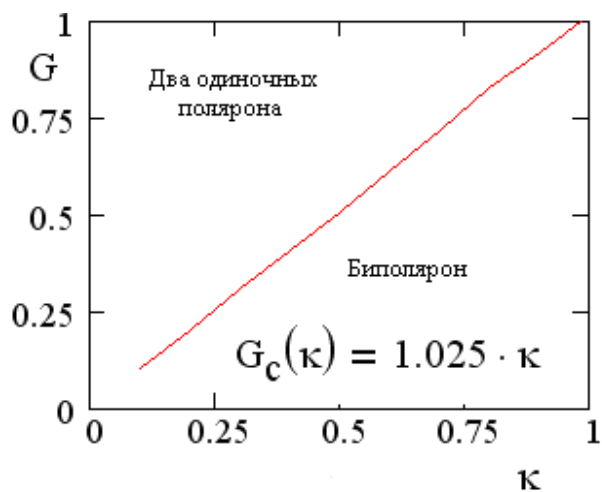

**Рис. 3** Фазовая диаграмма области существования биполярона для электронной ВФ (18a)

На рис. 3. представлена фазовая диаграмма существования континуального биполярона Холстейна, полученная нами для ВФ (18a). Фазовая диаграмма области существования БП – прямая линия. Обратим внимание, что дискретная модель [2], в отличие от наших расчётов, выполненных в рамках континуальной модели БП Холстейна, приводит к немонотонности линии, разграничивающей фазовые состояния.

Исследования теоремы вириала для полярона и для БП Холстейна показали, что никаких особенностей по сравнению с теоремой вириала для полярона и БП Пекара не наблюдается. Для полярона Холстейна выполняется теорема вириала 1:2:3:4 [8], для БП Холстейна, рассмотренного нами, выполняются те же вириальные соотношения, что и для БП сильной связи Пекара [7].

## ЗАКЛЮЧЕНИЕ

Применение континуального приближения, как показано в работе Холстейна [3], справедливо в том случае, когда координатная часть ВФ полярона изменяется плавно на расстояниях порядка постоянной решётки. В связи с тем, что функционал БП строился в виде суммы функционалов двух поляронов, в качестве критерия справедливости континуального приближения можно использовать критерий, полученный в [3] для изолированного полярона. В общем случае ВФ биполярона может распространяться на большее число постоянных решётки по сравнению с поляронной ВФ, однако, по порядку величины, критерий Холстейна не изменится. Так, согласно [3], линейный размер $L_p$ полярона Холстейна можно приближённо представить в виде:

$$L_p \approx a\left(\frac{4J}{A^2/M\omega_0^2}\right), \qquad (21)$$

где $a$ – расстояние между ближайшими соседями в линейной цепочке. Очевидно, что критерий существования полярона большого радиуса, когда $L_P \gg a$, выполняется при:

$$2J \gg A^2/2M\omega_0^2. \qquad (22)$$

Другими словами, ширина электронной зоны $2J$ велика по сравнению с величиной $A^2/2M\omega_0^2$. Обратное неравенство выполняется для полярона малого радиуса. На рис. 1 расстояние между ближайшими соседями в цепочке полагалось равным единице: $a = 1$. Видно, что в рассмотренном нами случае ВФ БП плавно меняется на расстояниях порядка постоянной решётки, поэтому критерий справедливости континуального приближения выполняется.

В работе [2] рассматривалась дискретная модель БП Холстейна–Хаббарда. По сравнению с континуальной моделью дискретная модель является точной, т.к. содержит в качестве своих предельных случаев и полярон большого радиуса, и полярон малого радиуса. При увеличении константы связи электрона с решёткой критерий выполнения континуального приближения перестаёт выполняться, а полярон переходит от полярона большого радиуса к полярону малого радиуса. Соответственно, при больших константах связи энергия полярона в дискретной цепочке не квадратична по константе связи [2]. В пределе небольших констант электрон-фононной связи, как показали наши расчёты, энергии связи биполярона, полученные в рамках дискретной и континуальной модели, хорошо согласуются между собой. При больших константах связи дискретная модель приводит к большим величинам энергии связи по сравнению с континуальной моделью.